# Research Challenges for Enterprise Cloud Computing

Ali Khajeh-Hosseini
Cloud Computing Co-laboratory
School of Computer Science
University of St Andrews
St Andrews, UK
akh@cs.st-andrews.ac.uk

Ian Sommerville
Cloud Computing Co-laboratory
School of Computer Science
University of St Andrews
St Andrews, UK
ifs@cs.st-andrews.ac.uk

Ilango Sriram
Department of Computer Science
University of Bristol
Bristol, UK
ilango@cs.bris.ac.uk

## ABSTRACT
Cloud computing represents a shift away from computing as a product that is purchased, to computing as a service that is delivered to consumers over the internet from large-scale data centers – or 'clouds'. This paper discusses some of the research challenges for cloud computing from an enterprise or organizational perspective, and puts them in context by reviewing the existing body of literature in cloud computing. Various research challenges relating to the following topics are discussed: the organizational changes brought about by cloud computing; the economic and organizational implications of its utility billing model; the security, legal and privacy issues that cloud computing raises. It is important to highlight these research challenges because cloud computing is not simply about a technological improvement of data centers but a fundamental change in how IT is provisioned and used. This type of research has the potential to influence wider adoption of cloud computing in enterprise, and in the consumer market too.

## Categories and Subject Descriptors
C.2.4 [**Computer-Communication Networks**]: Distributed Systems – *Cloud computing*

## 1. INTRODUCTION
Cloud computing has sparked a huge amount of interest in the IT industry. The market research and analysis firm *IDC* suggests that the market for cloud computing services was $16bn in 2008 and will rise to $42bn/year by 2012 [20]. Moreover, there has been a sharp rise in the number of cloud computing workshops and conferences recently and it is clear that academia is starting to take an interest in the new research challenges posed by cloud computing. There are many whitepapers and general introductions to cloud computing [e.g. 9, 44, 49, 52], and a number of authors have previously discussed research challenges in cloud computing [e.g. 1, 4, 36, 47, 54, 60]. However, these works have mostly focused on technical problems and little has been written about the research challenges for cloud computing from an enterprise or organizational perspective. This paper discusses some of these research challenges and puts them in context by reviewing the existing body of literature in cloud computing. We hope that this paper can act as a key resource for researchers in this fast moving field.

There seems to be many definitions of cloud computing [e.g. 1, 51, 60]. The US National Institute of Standards and Technology (NIST) has published a working definition [37] that seems to have captured the commonly agreed aspects of cloud computing. This definition describes cloud computing using:

- Five characteristics: on-demand self-service, broad network access, resource pooling, rapid elasticity, and measured service.
- Four deployment models: private clouds, community clouds, public clouds, and hybrid clouds.
- Three service models: Software as a Service (SaaS), Platform as a Service (PaaS), and Infrastructure as a Service (IaaS).

Large enterprises inevitably have a large number of computing systems that have been developed over a long period of time. These depend on different technologies, have different 'owners' within the enterprise and have complex dependencies both between the systems themselves, the data that they process, the middleware used and the platforms on which they run. Business processes have evolved to make use of the portfolio of systems available and these often rely on specific system features. Normally, there is no individual or group within the enterprise who knows about all of the systems that are in use and dependencies are often discovered by accident when something simply stops working after a change has been made. For international companies, different jurisdictions mean that the same system in different countries has to be used and supported in different ways.

Furthermore, IT provision is profoundly affected by political considerations. Senior management in the enterprise may set IT policies but these are left to individual parts of the enterprise to enact in their own way. Managers naturally tend to adopt strategies that benefit their part of the company and vice versa. The tension between central IT provision and end users has been constant since the 1960s with the constant complaint from users that central services are unwilling or unable to respond quickly to changing user requirements.

It is important to highlight cloud computing research challenges from an enterprise perspective because cloud computing is not simply about a technological improvement of data centers but a fundamental change in how IT is provisioned and used [13]. Enterprises need to consider the benefits, risks and the effects of cloud computing on their organizations and usage-practices in order to make decisions about its adoption and use. In the enterprise, the "adoption of cloud computing is as much dependent on the maturity of organizational and cultural (including legislative) processes as the technology, per se" [19].

Companies such as Amazon, Google and Microsoft have invested vast sums of money in building their public clouds and they seem to be leading the way in the technological innovation of clouds by releasing frequent updates and new features for their services. For example, a quick look at Amazon Web Services' (AWS) news archive[1] shows that they rolled-out over ten new and technologically impressive features to their cloud offerings in 2009. AWS also released a Security[2] and an Economic[3] center on their website, which shows that there is user demand for advice about the new economic and security implications of cloud computing. There is an opportunity for the academic community to address this demand by providing independent and impartial advice, tools and techniques to enterprise users who are interested in cloud computing.

Large organizations are inherently complex and for cloud computing to deliver real value to the enterprise rather than simply be a platform for simple tasks such as application testing or running product demos, the issues around migrating application systems to the cloud and satisfying the requirements of key system stakeholders have to be explored. These stakeholders include technical, project, operations and financial managers as well as the engineers who are going to be developing and supporting the individual systems. In enterprise, even desktop migration, which seems unproblematic from a cloud computing view, is challenging as we shall discuss later on. Small startup companies are unlikely to face many of the issues discussed in this paper as they have no legacy systems or existing business procedures; they can simply start building their systems in the cloud. From an enterprise perspective, costs are important but so too are customer relationships, public image, flexibility, business continuity and compliance. Enterprises need to understand how cloud computing affects all of these but this paper focuses on the following specific topics:

- The organizational change brought about with cloud computing (discussed in Section 2).
- The economic and organizational implications of the utility billing model in cloud computing (discussed in Section 3).
- The security, legal and privacy issues that cloud computing raises (discussed in Section 4).

Each section of this paper starts by reviewing the existing body of literature in cloud computing and commenting on the work from an enterprise perspective. A number of research challenges are then highlighted within each section and these challenges are further discussed in Section 5.

## 2. ORGANIZATIONAL CHANGE

In early 2008, Nicholas Carr's book titled *The Big Switch: Rewiring the World, from Edison to Google* was published [7], this was the first authoritative text to accessibly popularize cloud computing. In it, Carr pointed out that the current mode of delivery for most IT systems is similar in some respects to that of electricity, circa 1910. Back then, manufacturers had to build and maintain their own power generators even though that was not their main business. Today, most corporations build and maintain their own data centers even though that is not their main expertise. The result is a large number of "terribly inefficient" data centers [16]. Cloud computing could potentially solve this problem by enabling computing facilities such as storage, compute power, network infrastructure and applications to be delivered as a metered service over the internet, just like a utility.

However, currently the typical enterprise IT department is not used to a utility billing model across shared infrastructures; resource sharing across such infrastructures requires a certain level of cultural and organizational process maturity, and the move towards cloud computing will require significant changes to business processes and organizational boundaries [19]. Therefore, users need to consider the benefits, risks and the effects of cloud computing on their organizations and usage-practices in order to make decisions about its adoption and use: the potential for reduced costs could be just one of the persuasively significant benefits of cloud computing. Erdogmus [18] lists other benefits of cloud computing as "scalability, reliability, security, ease of deployment, and ease of management for customers, traded off against worries of trust, privacy, availability, performance, ownership, and supplier persistence". Some of these issues are discussed in subsequent sections of this paper.

Motahari-Nezhad *et al.* [40] briefly discussed the benefits and risks of using cloud computing from a business perspective. They highlighted the lack of environments for helping businesses migrate their legacy applications to the cloud. In addition, they pointed out the difficulties of finding and integrating different cloud services for a given set of business requirements. Motahari-Nezhad *et al.* proposed a conceptual architecture for a virtual business environment where individuals and SMEs can start and operate a virtual business using cloud-based services. This conceptual architecture includes four layers: business context, business services, business processes and IT services. Motahari-Nezhad *et al.* concluded by sketching an implementation of their conceptual architecture and the challenges encountered for each of the four layers. The work performed in the IT services layer of this conceptual architecture is going to be mostly affected by cloud computing.

Elson and Howell [17] provide just one example of the ways in which cloud computing could potentially affect the work of IT departments. They described how cloud computing can resolve conflicts in system development roles. For example, a start-up company that provides a hosting service for bloggers could, without cloud computing, be forced to build its own data center even though it only wants to integrate and offer existing open-source software to its customers. Elson and Howell described this scenario as conflating the roles of a software integrator and a hardware wrangler (someone who sets up and maintains hardware). They described how IaaS providers such as Amazon Web Services resolve such conflicts by providing an "explicit and narrow interface" in the form of a virtual machine (VM) image. Software integrators create the VM image and hardware wranglers deploy them.

---

[1] http://aws.amazon.com/about-aws/whats-new/2009/

[2] http://aws.amazon.com/security/

[3] http://aws.amazon.com/economics/

Yanosky [59] discussed how cloud computing will affect the authority of the IT department within universities; however, this also applies to IT departments in enterprise. The IT department gained its authority in the early days of computing when they had the majority of the programming skills and control of mainframes within an organization. As the use of IT expanded within organizations, system administrators and developers were forced to learn new skills as their role was no longer just about keeping the technology running. Until the invention of the PC, users relied on the services provided by the IT department for systems support.

The adoption of the PC eroded some of the IT department's authority as it provided users with an opportunity to create and use applications without the explicit support of the IT department. Users went on to form online communities to support each other as they were more experienced in solving technical problems. Although the IT department no longer had full control over the technology, it did have "a set of carrots and sticks at hand [...] including the supreme sanctions of refusing support for shadow systems [...] or cutting off network connectivity" [59].

The authority of the IT department is going to be further eroded by cloud computing. Cloud computing is increasingly turning "users into choosers" [59] who can replace the services provided by the IT department with service offered in the cloud. Yanosky mentioned that users will end-up asking for support from the IT department when they have problems in the cloud. However, this might not be the case as cloud service providers, such as Amazon, are starting to offer support services as well[4]. Yanosky suggested that IT departments could respond to this change by either controlling and monitoring the services that are allowed to be used in the cloud, or by providing a certification program, where they only support certified services. Either way, users will continue to have increased political influence within organizations, and the IT department's role will change from "provider to certifier, consultant and arbitrator" [59].

The type of organizational change that cloud computing results in can be demonstrated by taking a look at, for example, IT procurement within enterprise. Simplistically, procurement is based on obtaining estimates for things, then getting those estimates signed-off by management to allow the procurement to proceed. Capital and operational budgets are kept separate in this process, and it can take several months between the decision to procure hardware and the hardware being delivered, setup and ready to use. The use of cloud computing can greatly reduce this time period, but the more significant change relates to the empowerment of users and the diffusion of the IT department's authority as pointed out by Yanosky [59]. For example, a company's training coordinator who requires a few servers to run a week-long web-based training course can bypass their IT department and run the training course in the cloud. They could pay their cloud usage-bill using their personal credit card and charge back the amount as expenses to their employee. A similar scenario was recently reported by BP, where a group bypassed the company's procurement, IT department and security processes by using AWS to host a new customer facing website [32].

---

[4] http://aws.amazon.com/premiumsupport/

## 2.1 Research Challenges

The research done so far has provided some glimpses of the ways in which enterprise might change but an issue that has not been discussed in the literature to any significant extent is how to plan and design the actual organizational changes that will be required as IT provision changes as applications are migrated to the cloud, and as cloud services replace current desktop services. We know from 50 years of experience that the benefits of new technologies can only be realized when enterprises change their structure and processes to take advantage of technological innovation and, for sure, we can only realize the benefits of cloud computing if such changes take place.

There was a major organizational change when PCs became cheap enough to buy on individual departmental budgets and power shifted away from the IT department. Cloud computing is likely to result in a similar change but on a more significant scale, because power not only shifts away from the IT department, but also shifts outside the organization to companies such as Amazon. This raises a number of questions:

- What will be the changing role for the central IT department within organizations? Will their role change from "provider to certifier, consultant and arbitrator" as Yanosky [59] suggests, or will the complexity of IT systems and the lack of customized support from cloud providers and online support forums mean that organizations will still need central IT to provide and support most of their systems?

- How would compliance departments react to the migration of applications and data to cloud service providers? They might not have the same level of access to a cloud as they currently do to their internal systems, so how would they have to change their working practices?

- What are the political implications for organizations that lose control over some aspects of their services? Will it mean that moving to a cloud based system will be resisted by support personnel and system administrators who might either be worried about losing their jobs, or about no longer having complete control over a system? Would system administrators be happy to give up some of their control over systems and rely on cloud service providers for the support of end users?

- Would end users care about this? And would they change their working practices when central IT no longer has complete control over a system?

Currently when there is a problem with computer systems, organizations have coping strategies and workarounds that often rely on local expertise and knowledge. Users are usually good at knowing who has the local expertise about a particular system in their organization, and could go ask them for help or take them out for lunch when there is a problem. One of the reasons organizations might move services to the cloud is to reduce IT support costs so there will be less local expertise than before, and it is not clear who users could turn to once this expertise is lost.

Authority will change – the boss will no longer be able to demand that their problem gets priority from the support staff. In fact, the boss's authority may be diminished because they are not part of the organizational interface with the cloud provider. The diversity in current work environments and the ability to work while disconnected, allows coping strategies to be developed that may be impossible if everyone relies on a single networked provider for their applications. If the cloud goes down, does everything just stop? How will IT support departments need to change to cope with failures in the cloud?

We know from many years of ethnography that work practice in a setting evolves to reflect the systems and culture of that setting and that people develop work-arounds to cope with system problems and failures. How might current work-arounds change when the system is in the cloud rather than locally provided? Do the affordances of systems in the cloud differ from those that are locally provided? What 'cloud-based systems' (e.g. Twitter) might be used to support new kinds of work-arounds and communications?

It is often the case that, when people do things wrong, their mistakes are private and they can fix them before others find out about them. If there is a sharing model implemented in the cloud (rather than via some explicit action on the part of the sharers such as sending an email or uploading a file), then the mistakes made may become obvious immediately. Will people be aware of this and will they change their behavior?

## 3. COSTS

Cloud computing providers have detailed costing models and metrics that are used to bill users on a pay-per-use basis. This makes it easy for users to see the exact costs of running their applications in the cloud and it could well be that the design of their system can have a significant effect on its running costs. Armbrust *et al.* [1] mention short-term billing of as one of the novel features of cloud computing and a number of researchers have investigated the economic issues around cloud computing from a consumer and provider perspective.

### 3.1 Consumer Perspective

Youseff *et al.* [60] described the three pricing models that are used by cloud service providers, namely tiered pricing, per-unit pricing and subscription-based pricing. Tiered pricing is where different tiers each with different specifications (e.g. CPU and RAM) are provided at a cost per unit time. An example is Amazon EC2, where a small tier virtual machine has the equivalent of a 1.0GHz CPU, 1.7GB of RAM with 160GB of storage and costs $0.085 per hour at the time of writing, whereas as a large tier machine has the equivalent of four 1.0GHz CPUs, 7.5GB of RAM with 850GB of storage and costs $0.34 per hour[5]. Per-unit pricing is where the user pays for their exact resource usage, for example it costs $0.15 per GB per month to store data on Amazon's Simple Storage Service (S3)[6]. Subscription-based pricing is common in SaaS products such as Salesforce's Enterprise Edition CRM that costs $125 per user per month[7].

More elaborate pricing models exist in the grid and utility computing research community [6, 10, 50, 58] but Armbrust *et al.* [1] point to other utilities such as electricity and argue that simpler pricing models will remain dominant because they are transparent and understandable for users. However, new cloud computing pricing models based on market mechanisms are starting to emerge but it is not yet clear how such models can be effectively used by enterprise. An example of such models is used by Amazon's Spot Instances[8], which allows users to bid for unused capacity in Amazon's data centers. Amazon runs the user's instances as long as the bid price is higher than the spot price, which is set by Amazon based on their data center utilization.

Armbrust *et al.* [1] argue that elasticity – the ability to quickly scale up or down one's resource usage – is an important economic benefit of cloud computing as it transfers the costs of resource over-provisioning and the risks of under-provisioning to cloud providers. Armbrust *et al.* provide a few theoretical examples to highlight the importance of elasticity with respect to costs. An often cited real-world example of elasticity is Animoto.com whose active users grew from 25,000 to 250,000 in three days after they launched their application on Facebook [45].

The issue here for enterprises is that the majority of their systems have been built on the assumption that increases in demand will be supported by 'scaling up' to more powerful servers rather than 'scaling out' to larger numbers of servers, as Animoto did. Changing the architecture of these systems to support scaling out will inevitably be very expensive and in many cases will simply be impossible. Scaling out also raises some issues with respect to software licensing. For example, if application *A* relies on middleware *M*, typically the enterprise has negotiated the appropriate number of licenses for *M*. Running *A* on a cloud platform means that dynamic licensing for *M* may be required and this may be impossible to implement in practice. In general, the issues of licensing could be identified as a research challenge that has received some attention (e.g. [14]) but much more is required.

Klems *et al.* [31] attempted to address the problem of deciding whether deploying systems in the clouds makes economic sense. They highlighted some economic and technical issues that need to be considered when evaluating cloud adoption. The considerations were presented as a framework that could be used to compare the costs of using cloud computing with more conventional approaches, such as using in-house IT infrastructure. Their framework was very briefly evaluated using two case studies. However, no results were provided because the framework was at an early developmental stage and more conceptual than concrete.

---

[5] http://aws.amazon.com/ec2/#pricing

[6] http://aws.amazon.com/s3/#pricing

[7] http://www.salesforce.com/crm/editions-pricing.jsp

[8] http://aws.amazon.com/ec2/spot-instances/

Walker [55] also looked into the economics of cloud computing, and pointed out that lease-or-buy decisions have been researched in economics for more than 40 years. Walker used this insight to develop a model for comparing the cost of a CPU hour when it is purchased as part of a server cluster, with when it is leased (e.g. from Amazon EC2). The model was demonstrated using two scenarios, one where the cost of leasing was compared with purchasing a 60,000 core HPC cluster, and another where it was compared with purchasing a compute blade rack consisting of 176 cores. Walker showed that in both scenarios it would be cheaper to buy than lease when CPU utilization is very high (over 90%) and electricity is cheap. However, as expected, this would be reversed if CPU utilization is low or electricity is expensive. Walker's model was a good first step in developing models to aid decision makers, but it was too narrow in scope as it focused only on the cost of a CPU hour. Further work towards financial decision support is required that includes other costs such as housing the infrastructure, installation and maintenance, staff, storage and networking.

Assuncao *et al.* [2] investigated the use of clouds to extend the capacity of locally maintained clusters. They simulated the costs of using various strategies when borrowing resources from a cloud provider, and evaluated the benefits of doing this by using performance metrics such as the Average Weighted Response Time (AWRT) [22]. AWRT is the average time that user job-requests take to complete, shorter AWRTs means shorter waiting times. The investigation done by Assuncao *et al.* [2] is potentially useful for organizations that have private clouds and are looking to use public clouds when their in-house resources are over-utilised. However, cloud users expect instant resource allocation and AWRT might not be the best metric to measure performance improvement. Further research is required to identify the right metrics for evaluating the benefits of using public clouds to complement private clouds.

Deelman *et al.* [15] used simulation to calculate the cost of running a data-intensive scientific application on AWS. The focus of their study was to investigate the performance-cost tradeoffs of different execution plans by measuring execution times, amounts of data transferred to and from AWS, and the amount of storage used. They found the cost of running instances (i.e. CPU time) to be the dominant figure in the total cost of running their application. Others such as Kondo *et al.* [33] have found that the majority of the costs could be attributed to bandwidth usage. The study done by Deelman *et al.* [15] highlighted the potentials of using cloud computing as a cost-effective deployment option for data-intensive scientific applications. It assumed that the cost of running instances on AWS EC2 is calculated on a dollar-per-CPU-second basis (i.e. they normalized the costs). However, AWS charge on a dollar-per-CPU-hour basis and charge for a full hour even for partial hours. This means that launching 100 instances for 5 minutes would cost 100 CPU hours on AWS. This would result in significantly different costs from those shown by the simulations of Deelman *et al.* The most cost effective execution plans for AWS would be those that provision the right number of CPUs to ensure that jobs finish execution within full CPU hours and not partial hours.

## 3.2 Provider Perspective

From an enterprise perspective, in addition to investigating the costs of using cloud computing, it is important to investigate the costs of providing cloud computing services because:

1. For a range of security and legal reasons (discussed in Section 4), it is unlikely that all enterprise systems will be migrated to public clouds. Therefore, the use of private clouds will become important in enterprise. Creating private clouds is going to incur similar costs as current public clouds but probably on a smaller scale.

2. Once an enterprise private cloud is operational, the enterprise could potentially make additional profits by renting-out its spare IT capacity, just like a public cloud. Therefore, enterprise would benefit from the research carried out on the costs of cloud computing from a provider's perspective.

A few researchers have looked at the costs of cloud computing from a cloud provider's perspective. Greenberg *et al.* [21] discussed the costs of building cloud data centers and how they are different from building enterprise data centers. This topic was also addressed by Barroso and Hölzle [3] who went further and described the general challenges of building data centers that can be treated as a massive warehouse-scale computer (WSC). Greenberg *et al.* described how the costs of cloud data centers can be reduced by looking at the cost of servers, infrastructure, power, and networking. They presented some interesting ideas such as running data centers at hotter temperatures to reduce cooling costs and building micro data centers near users to reduce bandwidth costs. However, empirical research is required to investigate the actual benefits of such ideas. Barroso and Hölzle emphasized the importance of considering cost efficiency during the design of WSCs. They illustrated the significance of different cost parameters such as energy and server procurement by looking at a number of case studies that represented different deployment options. Barroso and Hölzle predicted that in the future, the costs of the data center facility (including energy usage) will become significantly larger than the actual server procurement costs.

Li *et al.* [34] also discussed the costs of cloud computing from a cloud provider's perspective. They developed a tool to model the Total Cost of Ownership (TCO) of setting up and maintaining a cloud by taking into account the costs of hardware, software, power, cooling, staff and real-estate. The tool would probably be useful for both large and small cloud providers as the cost factors are common to both. Calculating the TCO of a cloud starts by taking the required number of physical servers as an input. Lin *et al.* also provided a method for calculating a cloud's Utilization Cost. This allows cloud providers to calculate costs in a similar manner to TCO. However, rather than inputting the number of physical servers into the model, they can start by inputting the maximum number of virtual machines that they need in their cloud. The model would then calculate the required number of physical servers and racks depending on the virtual machine density that is defined by the user (this is the maximum number of virtual machines that can be deployed on a physical server). Modeling the Utilization Cost can be useful for cloud providers as it allows them to see and analyze a detailed breakdown of the total costs of building a cloud.

## 3.3 Research Challenges

In our view, the vast majority of current discussions on costs are simplistic when considered from an enterprise perspective. They simply consider the direct costs of computing provision, such as hardware and software costs and the costs of IT support. Few discussions of cost consider the economic issues around application migration. Nor do they consider existing procurement policies. How should these policies and processes be changed to accommodate the use of cloud computing in enterprise? And, if this is an interdisciplinary problem, how well is the field doing at collaborating with economists and business schools? They might already have answers to these questions, but the existing literature shows little signs of collaboration.

Currently, capital and operational budgets are often separately owned in many organizations, and procurement costs have to be known in advance before approval can be gained. Furthermore, specific signatories may be required to approve procurement and this mitigates against the use of 'on demand' systems. The utility billing model of cloud computing has a certain degree of uncertainty that goes against current procurement policies. The uncertainty relates to usage patterns and deployment options, and although some providers such as Google allow limits to be set on costs, this might not be enough. For example, an enterprise system deployed in the cloud that reaches its cost limit and stops working at a critical time is not going to be acceptable. Cost uncertainty is not regarded as an advantage in enterprise, so could cloud computing be compared with other utilities such as electricity and gas where usage patterns are mostly known and peaks can be anticipated? Will IT usage peaks need to be investigated to allow the costs of running systems in the cloud to be more accurately predicated?

Currently, cloud providers such as AWS do not have features to enable the allocation of costs to individual organizational units or projects, and Amazon's current model allows an AWS account to be used to purchase other goods and services from Amazon's shopping site. Some European businesses are asking AWS to post paper invoices to them[9]. These types of issues highlight the importance of auditability and compliance with financial regulations for enterprises who want to use cloud computing. However, these are all short term problems that can be addressed by cloud service providers, and the literature review provided in the previous sections show that academia is more interested in verifying the cost saving claims made by cloud providers. The research done so far has provided a good insight into the economics of cloud computing but further work is required to develop tools and techniques that allow users to examine the true costs of using cloud computing for their own purposes.

While costs are, of course, important for all enterprises, for individual managers, optimizing costs may not be their priority. Rather, their concerns are to ensure that they deliver what they have promised. The losses associated with failure to do so may far outweigh any costs savings from cloud migration and hence their priority is risk avoidance rather than cost saving. The main risks are mostly related to the security, legal and privacy issues that cloud computing raises; these are discussed in the next section.

---

[9] http://developer.amazonwebservices.com/connect/thread.jspa?threadID=24824

As an aside note, should enterprises be worried about moving towards a lease-only model of computing where a few big corporations provide computing resources? And what will be the regulatory issues that the authorities might have to deal with if there are only a small number of very big cloud providers? We should not forget that it used to be explicit IBM strategy to only lease equipment between the 1880s to 1950s because it was more profitable for them and less so for the customer. This was only stopped by anti-monopoly actions from the US government [12].

## 4. SECURITY, LEGAL AND PRIVACY ISSUES

Security, legal and privacy issues are widely acknowledged as being important in cloud computing. Most of the security and privacy issues in cloud computing are caused by users' lack of control over the physical infrastructure. This leads to legal issues that are affected by a cloud's physical location, which determines its jurisdiction. Furthermore, multitenancy brings the need for new solutions towards security and privacy. Jensen *et al.* [27] provided an overview of the technical security issues in cloud computing. Most of these issues are not specific to cloud computing as they relate to the underlying security problems of web services and web browsers. However, these security issues become more significant as cloud computing makes heavy use of web services and users rely on browsers to access services offered in the cloud.

For example, denial of service (DoS) attacks were a common concern even before cloud computing became popular, but when an application is targeted by a DoS attack in the cloud, the user or owner could actually end-up paying for the attack through their increased resource usage. This could be significantly higher than the peak usage of that application in an in-house data-center with limited resources. In fact, Jesper [28] recently reported this scenario with their application running on Amazon's AWS cloud but they did not mention if they were charged for the usage generated or if Amazon waived the extra costs incurred as a result of the attack.

Such incidents and other security concerns have resulted in the establishment of the Cloud Security Alliance, which is an industrial group with members from corporations such as Microsoft and HP. They have published a set of best practices and guidelines for organizations adopting cloud computing. These guidelines come in the form of problem statements and issues that need to be considered by cloud service consumers. They cover a wide range of areas including encryption and key management, portability and interoperability, and risk management [11]. The European Network and Information Security Agency (ENISA) has also published a report about security issues in cloud computing. They identified 35 risks of using cloud computing [8], which are split into the following categories:

- Policy and organizational risks such as vendor lock-in, loss of governance, compliance challenges, and cloud provider acquisition.

- Technical risks such as data leakage, distributed denial of service attacks, loss of encryption keys, and conflicts

- between customer hardening procedures and cloud platforms.
- Legal risks such as data protection and software licensing risks.
- Risks not specific to the cloud such as network problems, unauthorized access to data centers, and natural disasters.

In addition to the above risks, ENISA's report pointed out that security becomes cheaper to implement on a larger scale. Therefore cloud providers could potentially provide greater security measures such as hardening VM instances or hiring experts that can deal with particular threats [8].

Others have also considered the security benefits of using cloud computing. Armbrust *et al.* [1] discussed the security levels of systems deployed in local data centers and compared this with the potential security of using a cloud. In conclusion, they believe that systems deployed in the cloud could be made as secure as systems deployed in local data centers. They supported their hypothetical argument by mentioning the possible use of technologies such as encrypted storage to improve the security of a system deployed in the cloud. This work needs extension as it falls short of analyzing potential weaknesses of running systems in the cloud. One of the strengths of using SaaS applications in the cloud is that intruders do not have access to the application source code; this could be seen as a security advantage over non-SaaS applications where intruders have full access to the binary code [53]. Start-up companies could use cloud computing to eliminate the costs of developing a secure infrastructure. Kaufman [30] discussed security responsibility and asked if using cloud computing will result in security being a joint responsibility with the cloud providers. The question of responsibility was also raised in Mansfield-Devine [35], who highlighted the lack of control and checks over third parties who develop applications on top of the PaaS layer of the cloud.

Pearson [42] reviewed the main privacy risks in cloud computing. She summarized the risks from the perspective of individuals and organizations who wish to use cloud computing. From an individual's perspective, cloud computing presents risks of personal data exposure, and lack of awareness regarding the location and jurisdiction of their data. From an organization's perspective, cloud computing presents risks of legal liability, in addition to loss of reputation and user-trust in case of data loss. Pearson outlined a set of privacy requirements that need to be considered when developing applications or services using the cloud, including giving users notice of the information being collected and how it will be used, and allowing them to opt-out of their information being collected. The requirements were used to develop a number of design guidelines and "top tips" for software engineers developing cloud-based applications, these tips include minimizing the transfer of personal information to the cloud, and giving users control and choice over their data.

Joint *et al.* [29] provided an in-depth review of the legal issues that UK-based organizations should consider when using cloud computing. These include three of the eight principles of the UK's Data Protection Act (DPA), a law which came into place to protect personal data. For example, the first DPA principle requires organizations to ask for individual's permission before processing personal data. The use of cloud computing makes it difficult to provide individuals with full details of how and where their data will be processed. This lack of knowledge or transparency makes it difficult for individuals to reach "informed consent" [29]. In addition, Joint *et al.* pointed to legal issues regarding confidentiality, copyright, and specific rules that govern businesses that are regulated by the Financial Services Authority in the UK. Although Joint *et al.* highlighted the main legal issues that need to be considered, it is difficult to provide detailed guidelines as each cloud usage scenario is different.

Physical data location is important as there are no internationally agreed rules about data protection and privacy. For example, Amazon offer their cloud services from both U.S. and European-based data centers to be able to deal with differences in each region's rules and regulations. Jaeger *et al.* [26] reviewed a range of issues that are affected by the geographic location of a cloud, i.e. the location of the physical data centers. Jaeger *et al.* emphasized the importance of locality by pointing out that cloud computing increases the control of governments and corporations over resources. This is because cloud computing brings together vast amounts of data and computing resources in centralized data centers, compared to the current situation of hosting in geographically dispersed locations. While the location of the cloud has a significant effect on the rules and regulations that govern it, it is unclear "whether a cloud will be considered to legally be in one designated location [...] or in every location that has a data center that is part of the cloud" [26]. It is unlikely that these jurisdiction issues will stop the use of cloud services; however, they will have long-term implications that need to be considered by users.

Nelson [41] wrote a report as an advisory document for the US government. Nelson stressed the important role that governments can play in advancing the use of cloud computing. Just as procurement decisions made by the U.S. government in the 1980s were fundamental in the development of the Internet, governments have the power to advance cloud computing usage by being "model users" [41]. Nelson also provided advice to governments and policy makers on how to proceed; this included the need for governments to update their IT procurement rules to support procurement of systems with a cloud service model, as well as the need for them to encourage experimentation with the cloud. There are some signs that governments are taking on this type of advice, for example, the U.S. Federal Government has issued a Request for Information regarding IaaS offerings [23].

### 4.1 Research Challenges

From an enterprise perspective, security and regulatory issues are critical. Inevitably and understandably, most enterprises are likely to be cautious in moving their applications to the cloud simply because they do not really understand the security and regulatory issues involved. The difficulties in understanding the issues are exacerbated by the complexity of their systems – some data may have to be maintained within a specific jurisdiction, some data may be transferable to the cloud. Even where models of existing systems, their data and their dependencies exist, such information has simply not been required up till now and collecting such information is likely to be costly.

The work done so far has identified some of the security, legal and privacy issues that arise as a result of using cloud computing,

and this is an area where academia could be taking the lead. These are critical issues and their urgency is such that cloud users and their commercial providers might develop negotiated solutions in the immediate future. The legal issues could eventually be addressed by governments who might change existing regulations or introduce new ones. Some of the security issues such as DoS attacks are not specific to cloud computing but have new implications that need to be investigated. Such security breaches, and failures in general, seem harder to hide in the cloud as shown by the widely discussed Google Mail outages.

Security is about perceptions, many worry about security in the cloud because of a lack of control, while others argue that systems deployed in the cloud could be made as secure as systems deployed in local data centers [1]. Some organizations might be over confident about their internal security policies, but liability issues need to be resolved before enterprises start migrating applications to the cloud. The addition of third parties into the provider-consumer relationship introduces liability issues. For example, who is going to be liable for an incident that occurs while an enterprise 'bursts out' of their private cloud onto a public cloud? Regulations relating to data protection may place constraints on the movement of data and the national jurisdictions where it may be maintained. Furthermore, the cost and time to move data around could be a major bottleneck in the cloud. Enterprises with large volumes of data may therefore wish to specify where that data should be made available, when it may be moved around, etc. The practical issues that affect data migration in the cloud need to be investigated.

Another issue with data migration is compliance, which is especially challenging to satisfy when sensitive data is involved. For example, companies in the financial and health sectors have many regulatory requirements that restrict data movement. Compliance departments are likely to be conservative in their interpretation of the regulations and will require very detailed evidence that any movement of data outside the enterprise does not have associated compliance risks.

Some interesting issues arise when an organization wishes to control (in some way) the behavior of its staff using technological means. This may be a particular problem where the company uses a mixture of in-house systems and cloud-based systems. An example of an issue might be the inclusion of a link in a document to a site that is blocked by in-house systems but which can be accessed from documents in the cloud-based system. If this site distributes malware, there is then a security threat to the organization. How will companies react to this and will they be aware of the problems? How can individual users be made aware of the issues and what can be done if they do not follow the accepted rules? Of course, it may be that cloud providers will allow filters and rules to be applied to files that they manage. This might be fine if a company moves its services to the cloud provider as a whole but problems might arise when only some services are hosted on the cloud.

## 5. DISCUSSION

For sure, the cloud is a disruptive technology that will change the ways in which individuals and enterprises work. So, what does cloud computing mean for enterprises that have large-scale complex IT systems which are, in general, long-lifetime organizational, systems of systems? The difficult problems around such systems are:

- Understanding the requirements of diverse stakeholders, who are often uninterested in the system, until it is delivered and disrupts their lives.
- Understanding how processes and organizations have to change to make effective use of a system.
- So-called 'brownfield development' where new systems have to inter-operate with a range of existing systems [25].
- Designing for resilience so that essential services will be provided with a high level of availability [24].

So where does the cloud affect practice here? The issue of resilience is certainly one area – disaster recovery is potentially cheaper using external on-demand facilities. But how can the cloud contribute to other areas? Well, things that might help are:

- Easier prototyping and parallel operation. Systems can be installed on an external cloud for experiment then migrated to an internal server for operation. A common problem is that old and new systems cannot currently be run together because this requires hardware replication. Incremental development and delivery should be easier.
- Performance management. Issues of poor performance are common with new systems because the actual load is different from the predicted load. Obviously, the cloud makes it easier to provide extra facilities but it may also make it easier to do load simulations. However, sharing internal and external servers may introduce new problems because of the time required for data transfer to external systems.
- Trading off data and computation. Lots of problems with existing systems arise because the data models produced are rigid. If we have lots of available computational power, can we escape from this rigidity and design systems that have a much more flexible data model but which can maintain integrity by doing more computation? Would this be a real benefit however?
- Breaking down barriers created by current sandboxed IT systems. The cloud facilitates more mobile and more diffused organizations. Will using the cloud mean that the physical manifestations of many organizations are dramatically reduced (i.e. less office space)? Are there political issues here – might managers lose their visible empires? Does the cloud make it easier or harder to make work visible?

However, the problem with researching such organizational issues is that they are quite hard to investigate. The issues can certainly be flagged-up as we have done here in this paper, but some enterprises need to lead the way and take the plunge before these issues can be studied in real-life organizations. Making decisions about migrating parts of a large-scale complex IT system to the cloud is not that simple because a range of issues need to be considered, including:

- The effects of cloud computing on the work of IT departments and other parts of the organization.
- The costs, risks and benefits of running systems in the cloud.

These issues have a significant impact on migration decisions, and more research needs to be done in this area. It is important to note that the migration of IT systems to the cloud is not going to be done overnight – some predict that it could take between 10 to 15 years before enterprise makes this shift [48]. Therefore we are currently at the start of a transition period during which many decisions need to be made with respect to the migration of IT systems to the cloud. New tools and techniques need to be developed to help managers in enterprise make migration decisions.

One approach that could help decision making would be to develop a system modeling framework that can be used to model relevant attributes of large-scale IT systems. The framework would need a range of models such as models of application portfolios, models of data, cost models etc. These models could then be used to reason about and investigate migration decisions. For example, by modeling a system's hardware infrastructure and applications (at the executable level), it becomes possible to estimate the costs of running parts of that system in a cloud. However, costs alone are not sufficient to aid decision making – the risks and benefits of migration must also be included in the model. Furthermore, it would also be useful to include system stakeholders in the model as it would help identify the effects of migration on their work.

Extending existing modeling notations such as UML and SysML might be one way to achieve this. Relevant background for modeling applications and application portfolios includes work on modeling systems of systems [5, 57], enterprise system modeling [39, 43] and visualizing application portfolios [46]. Once applications are migrated to the cloud, the integration of legacy applications with cloud applications and the migration of data to the cloud must also be addressed [60].

Some enterprises have already started migrating desktop applications to the cloud [56]. Even though this seems to be one relevant use of cloud computing, it can be challenging to integrate such desktop applications with legacy systems. For example, some organizations use Microsoft Excel as a front-end to access legacy systems, which often do the actual data processing. It is unclear how the migration of such desktop applications, for example from Microsoft Office to Google Docs, is going to affect the integration of these applications with legacy systems.

In conclusion, this paper discussed various research challenges for cloud computing from an enterprise perspective, and put them in context by reviewing the relevant literature. Cloud computing as a research field is still in its early years. There seems to be an increasing number of workshops and conferences that are dedicated to the research challenges posed by cloud computing. It is encouraging to see such interest from the academic community and it is hoped that this paper can act as a good starting point for researchers interested in the research challenges for enterprise cloud computing.

The message here in our paper should not be taken as a negative comment on the use of cloud computing for enterprise, but that the research challenges highlighted here cannot be looked at from a purely technical perspective. These research challenges are interdisciplinary in nature, and there is a need for more co-operation between researchers, cloud users, and service providers. As previously mentioned, cloud computing is more than just a technological change and the issues that have been discussed in this paper have the potential to influence wider adoption of cloud computing.

## 6. ACKNOWLEDGMENTS

We thank the Scottish Informatics and Computer Science Alliance (SICSA) and Hewlett-Packard for funding the authors. We also thank the UK's EPSRC for funding the Large-Scale Complex IT Systems Initiative (www.lscits.org), which enabled our collaboration. We are grateful for the feedback provided by Dave Cliff, John Rooksby and Tim Storer on the drafts of this paper.